\documentclass{iopjournal}

\usepackage{amsmath}
\usepackage{bm}
\usepackage{algpseudocode}
\usepackage{algorithm}
\usepackage{algpseudocode}
\usepackage{amssymb}
\usepackage{xcolor}
\usepackage{hyperref}
\hypersetup{colorlinks = true, allcolors = blue}
\usepackage{cleveref}
\usepackage{subcaption}
\usepackage{soul}

\DeclareMathOperator*{\argmax}{arg\,max}

\DeclareMathOperator*{\Tr}{Tr}
\DeclareMathOperator*{\pT}{p_{\mathrm{T}}}

\DeclareMathOperator*{\MET}{{E}_{\mathrm{T}}^{\mathrm{miss}}}
\algnewcommand\algorithmicforeach{\textbf{for each}}
\algdef{S}[FOR]{ForEach}[1]{\algorithmicforeach\ #1\ \algorithmicdo}
\algdef{SE}[VARIABLES]{Variables}{EndVariables}
   {\algorithmicvariables}
   {\algorithmicend\ \algorithmicvariables}
\algnewcommand{\algorithmicvariables}{\textbf{predetermined variables}}
\algdef{SE}[REQUIRES]{Requires}{EndRequires}
   {\algorithmicrequires}
   {\algorithmicend\ \algorithmicrequires}
\algnewcommand{\algorithmicrequires}{\textbf{require}}

\begin{document}

\articletype{Article type} 

\title{Efficient Estimation of Unfactorizable Systematic Uncertainties}

\author{Alexis Romero$^{1,*}$\orcid{0000-0000-0000-0000}, Kyle Cranmer$^2$\orcid{0000-0000-0000-0000} and Daniel Whiteson$^{1}$\orcid{0000-0000-0000-0000}}

\affil{$^1$Department of Physics and Astronomy, University of California, Irvine CA}

\affil{$^2$UW Madison}

\affil{$^*$Author to whom any correspondence should be addressed.}

\email{alexir2@uci.edu}


\begin{abstract}
Accurate assessment of systematic uncertainties is an increasingly vital task in physics studies, where large, high-dimensional datasets, like those collected at the Large Hadron Collider, hold the key to new discoveries. Common approaches to assessing systematic uncertainties rely on simplifications, such as assuming that the impact of the various sources of uncertainty factorizes. In this paper, we provide realistic example scenarios in which this assumption fails. We introduce an algorithm that uses Gaussian process regression to estimate the impact of systematic uncertainties \textit{without} assuming factorization. The Gaussian process models are enhanced with derivative information, which increases the accuracy of the regression without increasing the number of samples. In addition, we present a novel sampling strategy based on Bayesian experimental design, which is shown to be more efficient than random and grid sampling in our example scenarios. 
\end{abstract}

\section{Introduction}

Experiments like those conducted at the Large Hadron Collider play a crucial role in precision tests of the Standard Model, where vast, high-dimensional datasets are collected for the precise measurement of particle properties. In many cases, the significance of a discrepancy between measurements and theoretical predictions relies crucially on the accurate estimation of systematic uncertainties. For example, recent measurements of the $W$ boson mass~\cite{CDF:2022hxs} and the muon anomalous magnetic moment~\cite{Muong-2:2021ojo} indicate discrepancies between measurements and predictions. These results are meaningful because they claim to be larger than their margins of systematic uncertainties. However, a full, rigorous assessment of the systematic uncertainties is often computationally intractable due to a generally large number of sources of uncertainty, which create a complex, high-dimensional space that must be characterized.

Several procedures have been established to deal with multiple sources of uncertainty in experiments. Most of them rely on simplifying assumptions, like the factorization of their underlying correlations. Examples of such procedures include orthogonalization~\cite{atlas2021jet,vos2016topquark,Aad_2015} and treatment of residual correlations~\cite{Aaboud_2019,Aaboud_2019_2}. These assumptions are often extended to treat the impact of individual sources of uncertainty on the detector response as also factorizable variables. For example, it is often assumed that the effect of individual sources (on-axis) can be used to estimate the combined effect of multiple sources (off-axis) of uncertainty.

In the following sections, we describe a strategy for characterizing the impact of systematic uncertainties without assuming factorization. Our strategy uses Gaussian processes (GPs) to regress observables of the detector response. Gaussian processes are flexible, nonparametric machine learning models that have been used in high-energy physics studies involving background selection and signal extraction~\cite{Frate:2017mai,Gandrakota_2023,bertone2016accelerating}. While GPs have been less widely applied to the estimation of systematic uncertainties~\cite{Heinrich:2019skt}, we show that they have the potential to outperform more traditional approaches. In addition, we show that we can leverage derivative information when fitting the GPs by computing the gradient of the observables with respect to the nuisance parameters associated to the sources of uncertainty. We show that GPs that are fit with this additional derivative information can efficiently estimate the complex and unfactorizable effects of multiple sources of systematic uncertainty.

In the example scenarios presented in this paper, the GP approach outperforms the traditional approach with significantly fewer samples. In more complex scenarios, however, more samples may be needed for the GP to provide a satisfactory estimate. In such cases, having an efficient strategy to sample the space of the detector response and obtain new training data is essential. 

We introduce a sampling strategy based on Bayesian Experimental Design (BED). This sampling strategy is designed to reduce the predictive uncertainty of a GP model as it selects new training data. 

Ideally, GP models should be both accurate, with low predictive error, and precise, with low predictive uncertainty. We show that in our example scenarios, the GP models conditioned using our BED sampling strategy have a lower overall mean-standard-error and predictive variance than those conditioned using random or grid sampling.

The rest of the paper is organized as follows: Sec.~\ref{sec:sys} describes the role of systematic uncertainties in statistical inference, and discusses common assumptions made in their assessment. Sec.~\ref{sec:GPR} describes key elements of GP regression, and sec.~\ref{sec:BED} introduces our BED strategy. In Sec.~\ref{sec:toy_1D}, we demonstrate our method on a one-dimensional toy example. Sections~\ref{sec:HEP_2D} and~\ref{sec:HEP_4D} extend this demonstration to more realistic high-energy physics scenarios in two and four dimensions, respectively. Sec.~\ref{sec:concl} contains our conclusions and future work.

\section{Systematic Uncertainties}
\label{sec:sys}

Experimental measurements in particle physics are typically based on the statistical inference of theoretical parameters of interest ($\bm{\theta}$) and nuisance parameters ($\bm{\nu}$) from experimental data ($\bm{n}$). The parameters of interest represent parameters of the theory, such as signal strength or resonance masses. By contrast, the nuisance parameters are treated as sources of systematic uncertainty. A major obstacle to inferring these parameters is the lack of access to a tractable likelihood, $p(\bm{n}|\bm{\theta}, \bm{\nu})$, which would make parameter estimation and limit-setting straightforward. Instead, the likelihood is often approximated as a product of Poisson terms for each bin of a histogram:

\begin{equation}
    p(\bm{n} | \bm{\theta}, \bm{\nu}) \approx \prod \limits_{i \in \mathrm{bins}} \mathrm{Pois}
 (n_i | \eta_i(\bm{\theta}, \bm{\nu})), i=1, \ldots, B,
\end{equation}
where $\bm{n}=(n_1, \ldots, n_B)$ are the observed bin counts and $\eta_i(\bm{\theta}, \bm{\nu})$ the number of expected events in bin $i$, estimated from simulation. 

The expected number of events is often expressed as a function of the parameters, 
\begin{equation}
    \eta_i(\bm{\theta}, \bm{\nu}) = s_i(\bm{\theta}) \epsilon_{s,i}(\bm{\nu}) + b_i \epsilon_{b,i}(\bm{\nu}),
\end{equation}
where $s_i(\bm{\theta})$ and $b_i$ are the expected number of signal and background events, respectively, and $\epsilon_{s,i}(\bm{\nu})$ and $\epsilon_{b,i}(\bm{\nu})$ their efficiencies.

In addition, some auxiliary measurements $\bm{a}$ may be used to constrain the nuisance parameters through additional prior estimates\cite{cranmer2015practical}, $p_{\mathrm{c}}(\bm{a} | \bm{\nu})$. If there are multiple nuisance parameters, they are typically assumed to be independent, and thus their contributions to the model are assumed to be the product of the individual parameters. All in all, the likelihood can be approximated as
\begin{align}
    p(\bm{n}, \bm{a} | \bm{\theta}, \bm{\nu}) \approx & \prod \limits_{i \in \mathrm{bins}} \mathrm{Pois}
 (n_i | \eta_i(\bm{\theta}, \bm{\nu}) ) \cdot \prod \limits_{\nu_j \in \bm{\nu}} p_{\mathrm{c}}(\bm{a} | \nu_j),
 \label{eq:priorx_theta_nu}
 \end{align}
However, even if the nuisance parameters are independent such that the product of $p_{\mathrm{c}}(\bm{a} | \nu_j)$ holds, that does not guarantee that the experimental response (i.e. $\epsilon_{s,i}(\mathbf{\nu})$ and $\epsilon_{b,i}(\mathbf{\nu})$) can be factorized in terms of the individual parameters. An example of this is the response of the efficiencies to simultaneous changes in the nuisance parameters, which will likely not factorize.


\section{Gaussian Process Regression}
\label{sec:GPR}

A GP~\cite{Frate:2017mai} is a collection of random variables, such that the joint distribution of any finite choice of variables forms a multivariate Gaussian. To illustrate this, consider a set $\mathcal{D}$ of $N$ observation pairs, {$\mathcal{D}=\{(\bm{x}_i, y_i) |  i=1,\ldots,N\}$}, where $\bm{x}_i \in \mathbb{R}^D$ denotes an input vector and $y_i \in \mathbb{R}$ denotes a scalar output. In the GP framework, the output values $y_i$ are seen as being drawn from a Gaussian distribution with mean function $\mu(\bm{x})$ and covariance function $k(\bm{x}, \bm{x}')$. The Gaussian process of a real function $f(\bm{x})$ can be defined as 
\begin{align}
    f(\bm{x}) & \sim 
    {\mathcal{N}}
    (\mu(\bm{x}), k (\bm{x}, \bm{x}')), \\
    \mu(\bm{x}) & = \mathbb{E} [ f(\bm{x}) ] \\
    k (\bm{x}, \bm{x}') & = \mathbb{E} [ (f(\bm{x}) - \mu(\bm{x})) (f(\bm{x}') - \mu(\bm{x}'))].
\end{align}
For simplicity, we assume the mean function to be zero, although this may not be necessary. 

The covariance function specifies the covariance between any two pairs of variables. We utilize the \textit{squared exponential} (SE) function, also known as the radial basis function, as the covariance function:
\begin{equation}
    \mathrm{cov} \left[ y_i, y_j \right] = k \left( \bm{x}_i, \bm{x}_j \right) =
    \alpha^2 \exp \left( -\frac{\left\| \bm{x}_i-\bm{x}_j \right\|^2}{2 {\bm{\ell}}^2} \right),
    \label{eq:k_yy}
\end{equation}
where the parameters $\alpha \in \mathbb{R}$ and $\bm{\ell} \in \mathbb{R}^D$ control the amplitude and length scale of the squared exponential, respectively. 

To simplify the notation, we will aggregate the input vectors into the matrix $X$ of size $N \times D$, and the output targets into the vector $\bm{y}$ of length $N$, such that $\mathcal{D}=\{(X, \bm{y}\})$. We can incorporate the knowledge that these observations provide by using them to condition the joint prior Gaussian distribution--a process also known as training. The conditioned GP model can then be used to predict new output $\bm{y}_*$ at test input $X_*$:
\begin{align}
    (\bm{y}_* &| X_*, X, \bm{y}) \sim \mathcal{N} (\bm{\mu}_*, \Sigma_*), 
    \\
    \bm{\mu}_* &= K \left( X_*, X \right) [K \left( X, X \right) + \sigma^2 I]^{-1} \bm{y}, 
    \label{eq:mu_gpreg_noisy}
    \\
    \begin{split}
    {\Sigma}_* &=
    K \left( X_*, X_*\right) - K \left( X_*, X \right) [K \left( X, X \right) + \sigma^2 I]^{-1} K \left( X, X_* \right).
    \label{eq:sigma_gpreg_noisy}
     \end{split}
\end{align}
where $\bm{\mu}_*$ and ${\Sigma}_*$ are the mean and covariance matrix of the joint posterior distribution, respectively. If there are $N_*$ test inputs, $K \left( X, X_* \right)$ denotes $N \times N_*$ matrix of covariances evaluated at each pair of training and testing inputs, and similarly for the matrices of covariances $K \left( X, X \right)$, $K \left( X_*, X_* \right)$, and $K \left( X_*, X \right)$. The $\sigma^2 I$ term represents additive white Gaussian noise with standard deviation $\sigma^2$. This term is typically added to approximate statistical fluctuations in the model and the input. 

\subsection{Including Derivative Information}
\label{sec:GPRwDerivs}

If derivative observations of the function $f(\bm{x})$ are available, they can be incorporated into the GP model. Suppose that we are given a new set of observations corresponding to the $N$ first partial derivatives of $f(\bm{x})$ along dimension $d$: $\mathcal{D}_d = \{(\bm{x}_{i}, w_{d, i})\}$, $d = 1, \ldots, D$, $i=1, \ldots, N$, where 
\begin{equation}
    w_{d, i} = \frac{\partial f (\bm{x}_i)}{\partial x_d}.
\end{equation}

The covariances between function and derivative observations must satisfy~\cite{rasmussen:2008}
\begin{align}
    \mathrm{cov} \left[ w_{d, i}, y_{j} \right]
    &= \frac{\partial}{\partial x_d} \mathrm{cov} \left[ (y_{i}, y_{j}) \right],
    \label{eq:fun_derivative_cov}
    \\
    \mathrm{cov} \left[ w_{d, i}, w_{e, j} \right]
    &= \frac{\partial^2}{\partial x_d \partial x_e} \mathrm{cov} \left[ (y_{i}, y_{j}) \right].
    \label{eq:derivative_cov}
\end{align}

Conditioning and inference are evaluated as usual, following the joint posterior distributions in Eq.~\ref{eq:mu_gpreg_noisy} and~\ref{eq:sigma_gpreg_noisy}, but the matrix of inputs $X$, and the output vector $\bm{y}$ must be expanded to include derivative observations. Likewise, the covariance matrices are expanded to include the covariances between all pairs of function and derivative observations according to Eq.~\ref{eq:fun_derivative_cov} and~\ref{eq:derivative_cov}.

The statistical fluctuations of the derivative observations may be different from those of the function observations. To account for these fluctuations, we add white Gaussian noise with standard deviation $\sigma_{\mathrm{der}}^2$ to the elements of the covariance matrix that are conditioned on derivative observations.

Throughout this paper, we refer to a GP model that incorporates derivative observations as a \textit{derivative} GP. By contrast, we refer to a GP model without derivative observations as a \textit{regular} GP.

\subsection{Efficient Gaussian Process Regression with Approximated Gradients}

The physics examples in this paper focus on estimating the efficiency $\epsilon$ as a function of the nuisance parameters, $\bm{\nu}$. For large datasets, estimating the efficiency is generally a computationally expensive task involving evaluating each event in the dataset over predetermined selection criteria.

Selection criteria in high-energy physics analyses are often based on statistical cuts, which are non-differentiable. To obtain the gradients of the efficiency with respect to the nuisance parameters, we replace these hard cuts with sigmoid functions ($S$), which are easily differentiable. The sigmoid functions are modified with the parameters $a$ and $c$, controlling the steepness and horizontal shift of the curve, respectively:
\begin{equation}
    S(x, a, c) = \frac{1}{1 + e^{-a(x-c)}}.
    \label{eq:sigmoid}
\end{equation}
{For example, to estimate the fraction of events with $\pT < 200$ GeV, one could simply count the number of events that fall below this threshold. Alternatively, this could be approximated using the modified sigmoid function with $c=200$ and a negative value for $a$. The magnitude of $a$ is tuned according to the specific application. Absolute values smaller than unity result in more elongated curves, while those larger than unity result in steeper curves that more closely resemble hard cuts.}

Methods using numerical differentiation can also be used to approximate the gradients of non-differentiable detector observables~\cite{Pinder2022,Simpson_2023}. These numerical gradients can enhance GP models without the need for analytical differentiation. In our studies, however, we find that the simple approach of substituting numerical cuts with sigmoid functions, and computing their gradients analytically, results in more accurate GP models than those conditioned on numerical gradients.

Lastly, we note that derivative GPs have a higher computational cost than regular GPs--$\mathcal{O}(N^3 D^3)$ vs. $\mathcal{O}(N^3)$ when conditioned on $N$ samples in $D$ dimensions. However, we expect the training and inference time of the GPs to be relatively negligible compared to the time it may take to sample the efficiency or other observables of the detector response. In addition, recent work has been done to scale derivative GPs in low- \cite{Eriksson2018ScalingGP} and high-dimensions~\cite{deroos2021highdimensional,padidar2021scaling}.

\section{Bayesian Experimental Design}
\label{sec:BED}

We introduce a BED strategy designed to efficiently select new training observations while reducing the predictive uncertainty of a GP model. At the core of the BED strategy is the utility function, $U$. 

\subsection{Utility}
\label{sec:utility}

Suppose we have collected $N$ input/output pairs, $\mathcal{D} = \{(\bm{x}_i, y_i) | $i=1, \ldots, N$\}$, which have been used to train a GP model. We then wish to make an additional observation, but due to high sampling costs, our goal is to make an informed decision and find the $(\bm{x}_{N+1}, y_{N+1})$ pair that is expected to reduce the predictive uncertainty of the model the most. {We design the utility function to guide us in finding this new observation.}

{The computation of the utility can be intuitively described as follows: We begin by selecting a set of inputs, which we call the \textit{utility input} (${X}_{u}$), and are intended to be representative of the input space. A good choice of utility input could be, for example, a set of grid points that uniformly span the input domain. Using this set, we compute the posterior covariance matrix ${\Sigma}_{u}$, where each entry corresponds to a pair of points in ${X}_{u}$, according to Eq.~\ref{eq:sigma_gpreg_noisy}.}

{For some arbitrary new input $\bm{x}$, we would like to estimate the loss in the predictive uncertainty of the model if we were to include $\bm{x}$ in the training set. To do this, we temporarily augment the model with the fictitious observation $(\bm{x}, \mu)$, where $\mu$ is the predictive mean at $\bm{x}$, as given by Eq.~\ref{eq:mu_gpreg_noisy}. With this augmented model, we recalculate the posterior covariance matrix ${\Sigma}_{u}^{\mathrm{aug}}$ on the utility input.
 By comparing ${\Sigma}_{u}$ and ${\Sigma}_{u}^{\mathrm{aug}}$, we can quantify the relative loss in the predictive uncertainty that would result from sampling at $\bm{x}$.}

Many quantities could be used to characterize the predictive uncertainty of a model given the posterior covariance matrix, such as:
\begin{itemize}
    \item \textbf{Determinant}: Considers the determinant of the posterior covariance matrix, which is correlated to the error ellipsoid of the model.  
    \item \textbf{Trace}: Considers the trace of the posterior covariance matrix, which quantifies the net variance of the model.
    \item \textbf{Maximum variance}: Considers the maximum entry in the diagonal of the predictive covariance matrix, which quantifies the maximum variance of the model.
\end{itemize}
In our studies, we find that using the trace is a good and computationally efficient way to reduce the predictive variance in the GP models, thus reducing the overall uncertainty. In addition, the trace proves to be more numerically stable than the determinant, which can tend to zero when the input variables are strongly correlated.

Ideally, the trace of ${\Sigma}_{u}^{\mathrm{aug}}$ should be less than the trace of ${\Sigma_{u}}$. In light of this, we define the utility as 
\begin{equation}
    \mathrm{U}(\bm{x}, \mathcal{D}, X_u) \leftarrow 1 - \Tr({\Sigma}_{u}^{\mathrm{aug}}) / \Tr({\Sigma_{u}}). 
    \label{eq:utility}
\end{equation}
The pseudocode of the utility function is shown in Algorithm~\ref{alg:utilityfun}. While it is also possible to define the utility as simply the negative of ${\Sigma}_{u}^{\mathrm{aug}}$ with the same effect, we prefer the form above because it more clearly reflects the relative loss in the variance of the model. 

\algrenewcommand\algorithmicrequire{\textbf{Input:}}

\begin{algorithm}
\caption{Utility function (U)}
\label{alg:utilityfun}
\begin{algorithmic}
\Require $(\bm{x}, \mathcal{D}, {X}_u)$
\State Evaluate $\Sigma_u$ at every point in ${X}_u$ according to Eq.~\ref{eq:sigma_gpreg_noisy}.
\State Evaluate $\mu$ at $\bm{x}$ according to Eq.~\ref{eq:mu_gpreg_noisy}.
\State Temporarily augment the training dataset: \\ ~~ $\mathcal{D}^{\mathrm{aug}} \leftarrow \mathcal{D} \cup \{(\bm{x}, \mu)\}$.
\State Temporarily recondition the model on $\mathcal{D}^{\mathrm{aug}}$.
\State With the augmented model, evaluate $\Sigma_u^{\mathrm{aug}}$ at every point in ${X}_{u}$ according to Eq.~\ref{eq:sigma_gpreg_noisy}.
\State \Return $1 - \Tr({\Sigma}_u^{\mathrm{aug}}) / \Tr({\Sigma}_{u})$.
\end{algorithmic}
\end{algorithm}

\subsection{Sensitivity to the Utility Input $X_u$}

The BED strategy is sensitive to the choice of the utility input, $X_u$. Ideally, evaluating the model at this input utility should provide a sense of the quality of the model's predictions over the input space. Thus, high-resolution grids with densely spaced input points are good candidates for $X_u$. However, this choice has two disadvantages: First, evaluating the model at each point in $X_u$ can quickly become computationally expensive, particularly for models with high dimensionality. Second, it does not guarantee that the BED would not overtrain on $X_u$ by repeatedly selecting new samples that lie on or near the grid points.

To address the second issue mentioned above, we add regularization to $X_u$. After each BED iteration, every point in $X_u$ is perturbed by adding Gaussian noise with $\gamma^2$ standard deviation. {This stochastic perturbation helps reduce the risk of overfitting by encouraging diversity in the evaluation points.} In addition, choosing an adequate $\gamma^2$ parameter may also improve the performance of the BED, even when using low-resolution grids for $X_u$. See Sections~\ref{sec:HEP2d_BED} and~\ref{sec:HEP4d_BED} for further discussion.

\subsection{Updating the Beliefs About the Model}

At each BED iteration, the utility function is maximized, and a new sampling input is selected: 
\begin{equation}
    \bm{x}_{N+1} = \argmax_{\bm{x}} \mathrm{U}(\bm{x}, \mathcal{D}, X_u).
\end{equation}
Subsequently, we sample $y_{N+1}$ and recondition the GP model with the updated set of input/output observations, $\mathcal{D} \leftarrow \mathcal{D} \cup \{(\bm{x}_{N+1}, y_{N+1})\}$. This process is repeated until reaching a stopping criterion. 

\section{\label{sec:toy_1D} Simple 1D Toy Model}

\begin{figure}[!htb]
\centering
\begin{subfigure}{.45\textwidth}
    \centering
    \includegraphics[width=0.9\linewidth]{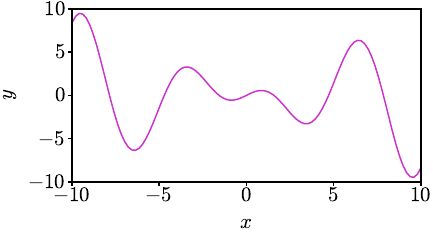}
\end{subfigure}
\begin{subfigure}{.45\textwidth}
    \centering
    \includegraphics[width=0.9\linewidth]{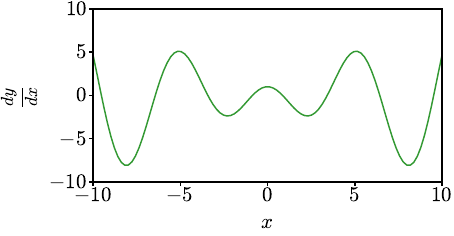}
\end{subfigure}
\caption{Function ($y$) and and derivative ($\frac{dy}{dx}$) output values of the simple 1D toy model for $x \in [-10, 10]$.}
\label{fig:toy1d_truth}
\end{figure}

In this section, we test the GP regression and BED strategy on a simple 1D toy model that is described by the following sinusoidal dynamical system with scalar input $x$ and output $y$:
\begin{align}
    y &= x \cos(x), \\ 
    \frac{dy}{dx} &= \cos(x) - x \sin(x),
\end{align}
The function and its derivative are shown in Fig.~\ref{fig:toy1d_truth}.

\subsection{\label{sec:1D_GPR} Gaussian Process Regression}

\begin{figure}[!htb]
\centering
\begin{subfigure}{.45\textwidth}
  \centering
  \includegraphics[width=0.9\linewidth]{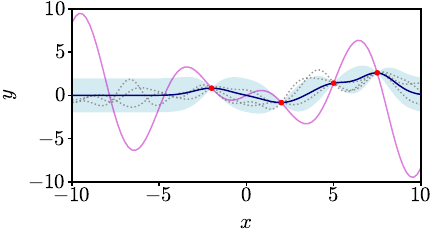}
  \caption{Regular GP}
  \label{fig:1D_regGP_init}
\end{subfigure}%
\begin{subfigure}{.45\textwidth}
  \centering
  \includegraphics[width=0.9\linewidth]{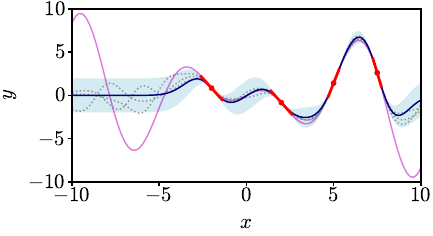}
  \caption{Derivative GP}
  \label{fig:1D_derGP_init}
\end{subfigure}
\caption{Panel (a) shows the four training observations (red dots) used to fit the regular GP model. Also shown are the predicted mean $\mu_{\mathrm{test}}$ (blue line) and confidence band (light blue area), together with three functions sampled from the posterior (gray dotted lines). {For reference, the true underlying function is depicted in magenta.}
Similarly, panel (b) shows the predictions of the derivative GP model using the same four observations but augmented by their gradients (red tangent segments).}
\label{fig:toy1d_initial_GPR}
\end{figure}

To train the GP, we sample the system at four random points within $x \in [-10, 10]$. These four observations are used to train a noise-free regular GP model with hyperparameters $\alpha=1$ and $\ell=1$. The models are tested on 100 points uniformly distributed within the same range. The mean prediction ($\mu_{\mathrm{test}}$) and the uncertainty bands are shown in Fig.~\ref{fig:toy1d_initial_GPR}. 

Using the same hyperparameters and observations (plus their gradients), we also train a noise-free derivative GP. The results are shown in Fig.~\ref{fig:toy1d_initial_GPR}.

The regular and derivative GPs reveal different predictions even when trained on the same observations. By incorporating gradient information, the derivative GP is able to more accurately predict the turning points near the training observations. On the other hand, the regular GP struggles to identify most turning points, as well as the inflection point at $x=0$.

\subsection{\label{sec:1D_BED} Bayesian Experimental Design}

{We evaluate the BED strategy by performing 20 iterations of the algorithm.} After each iteration, the predictive mean $(\mu_{\mathrm{test}})$ and the predictive covariance matrix $(\Sigma_{\mathrm{test}})$ of the models are updated. We evaluate the quality of the predictions by calculating the mean-squared-error (MSE), which provides a sense of accuracy. We also calculate the trace of the covariance matrix, $\mathrm{Tr}(\Sigma_{\mathrm{test}})$, which is proportional to the variance of the model, providing a sense of precision. Lower $\mathrm{Tr}(\Sigma_{\mathrm{test}})$ values suggest lower predictive uncertainty.

\begin{figure}[!htb]
\centering
\begin{subfigure}{.45\textwidth}
  \centering
  \includegraphics[width=0.9\linewidth]{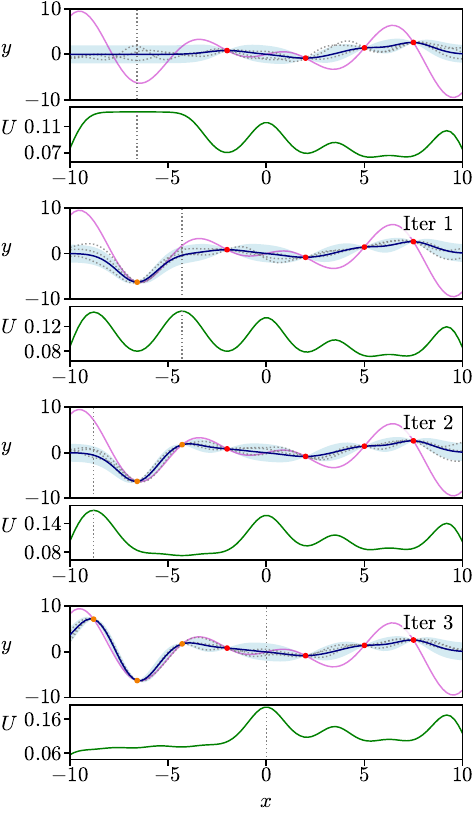}
  \caption{Regular GP}
  \label{fig:1D_regGP_BED}
\end{subfigure}%
\begin{subfigure}{.45\textwidth}
  \centering
  \includegraphics[width=0.9\linewidth]{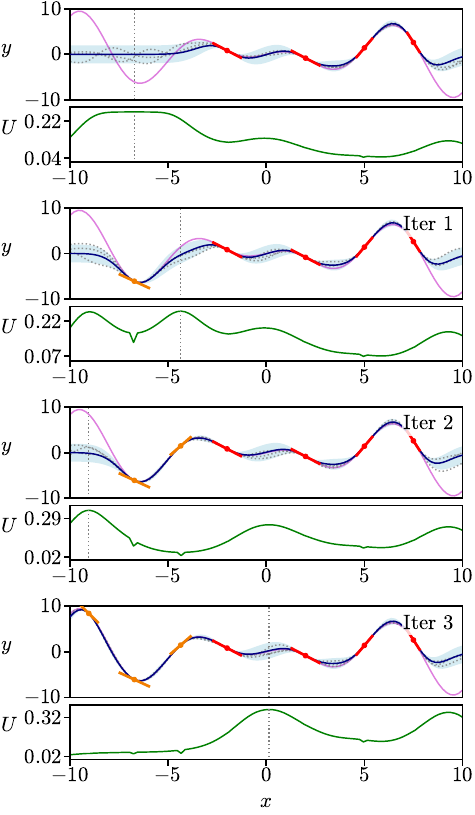}
  \caption{Derivative GP}
  \label{fig:1D_derGP_BED}
\end{subfigure}
\caption{First three BED iterations of the regular (a) derivative (b) GP models. The panels show the initial observations (red dots) and the observations selected by the BED strategy (orange dots). Also shown are the predictive mean $\mu_{\mathrm{test}}$ (blue line), the uncertainty band (light blue area), and three functions sampled from the posterior (gray dotted lines). {For reference, the true underlying function is depicted in magenta. Along the bottom of each panel, we show the utility function in green. The location of the next observation is indicated by the dotted vertical line, corresponding to the point that maximizes the utility.}}
\label{fig:toy1d_BED}
\end{figure}

This strategy requires specifying the utility input $X_{u}$. In this example, we choose it to match the test set: 100 points uniformly distributed in the range $x \in [-10, 10]$. Fig.~\ref{fig:toy1d_BED} illustrates how the GP models evolve over the first three BED iterations. With every iteration, the utility is maximized, and a new training observation is selected. As we recondition the models to include the latest observations, their predictions resemble more and more the function shown in Fig.~\ref{fig:toy1d_truth}, while the uncertainty band decreases in area.

\begin{figure}[!htb]
    \centering
    \includegraphics[width=0.75\textwidth]{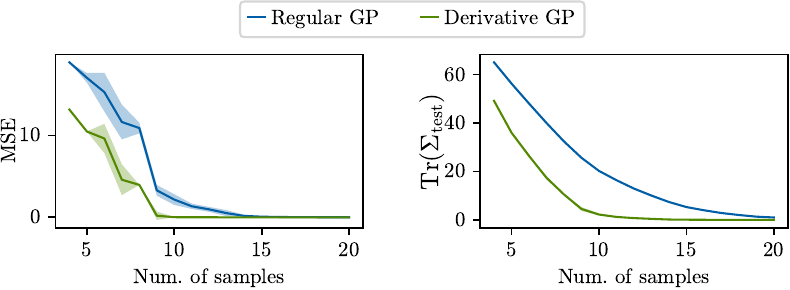}
    \caption{MSE and $\mathrm{Tr}(\Sigma_{\mathrm{test}})$ of the models after each of the 20 BED iterations. The solid lines represent the mean values of 10 calls of the BED strategy with different initial random seeds, and the shaded areas represent the standard deviation.}
    \label{fig:toy1D_BED_mse_uncert}
\end{figure}

To test the sensitivity of the BED strategy to the initial random conditions, the algorithm is called ten times using different initial random seeds. The results are shown in Fig.~\ref{fig:toy1D_BED_mse_uncert}. The derivative GP outperforms the regular GP with lower MSE and $\mathrm{Tr}(\Sigma_{\mathrm{test}})$ values, corroborating the power of incorporating gradients into the GP model.

\section{\label{sec:HEP_2D}High-Energy Physics Experiments: 2D Efficiency Estimation}

In this section, we apply GP regression and the BED strategy to experiments in high-energy physics. We focus on estimating the efficiency ($\epsilon$) as a function of two nuisance parameters, the jet energy scales. Efficiency often plays a vital role in estimating systematic uncertainties (Eq.~\ref{eq:priorx_theta_nu}), and having a strategy that can accurately estimate the efficiency based on limited samples can be a powerful tool in high-energy physics experiments.

\begin{figure}[!htb]
    \centering
    \includegraphics[width=0.38\textwidth]{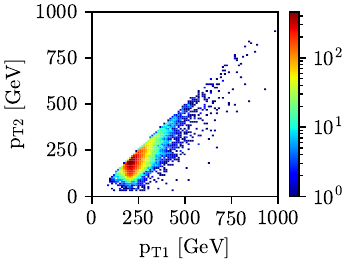}
    \caption{Distribution of the $\pT$ of the two leading jets.}
    \label{fig:pT_dist}
\end{figure}

The dataset consists of 30K events, each with three jets. The jets are arranged in descending order according to their transverse momentum ($\pT$). The hardest jet in an event is labeled as $j_1$, followed by $j_2$ and $j_3$. The $\pT$ distribution of the two leading jets is shown in Fig.~\ref{fig:pT_dist}.

We consider the following jet energy scales:
\begin{itemize}
    \item $\nu_{1}$: The jet energy scale of the leading jet, $j_1$.
    \item $\nu_{2,3}$: The jet energy scale of the two softer jets, $j_2$ and $j_3$.
\end{itemize}
And the following $\pT$-based selection criteria: 
\begin{equation}
\begin{split}
    \frac{\pT_{1}}{\nu_{1}}  &> 200 \mathrm{~GeV},\\
    \frac{\pT_{2}}{\nu_{2,3}} &< 200 \mathrm{~GeV},
\end{split}
\label{eq:sc}
\end{equation}
where $\pT_{1}$ and $\pT_{2}$ are the nominal transverse momenta of $j_1$ and $j_2$, respectively. {Let $\mathrm{N_{pass}(\nu_1, \nu_{2,3})}$ be the number of events that satisfy the selection criteria. From this selected subset, we calculate the number of events with a missing transverse energy ($\MET$) of less than 50 GeV, denoted as $\mathrm{N_{\MET < 50 GeV}(\nu_{1}, \nu_{2,3})}$. The efficiency is defined as the ratio of these two quantities}:
\begin{equation}
    \epsilon(\nu_{1}, \nu_{2,3}) = \frac{\mathrm{N_{\MET < 50 GeV}}(\nu_{1}, \nu_{2,3})}{\mathrm{N_{pass}(\nu_1, \nu_{2,3})}}. \label{eq:effcut}
\end{equation}
And the missing transverse energy is calculated according to the following formula:
\begin{equation}
    \MET(\nu_{1}, \nu_{2,3}) = \sqrt{ \left( \sum_{i=1, 2, 3} \frac{\pT_{ix}}{\nu_{i}} \right)^2 + \left( \sum_{i=1, 2, 3} \frac{\pT_{iy}}{\nu_{i}} \right)^2}.
    \label{eq:MET}
\end{equation}

\begin{figure}[!htb]
    \centering
    \includegraphics[width=0.38\textwidth]{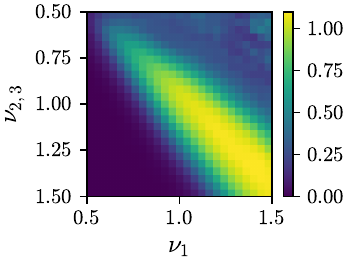}
    \caption{Normalized efficiency as a function of the jet energy scales.  The image is composed of $25 \times 25$ ``pixels" of size $0.4 \times 0.4$ in $(\nu_1, \nu_{2,3})$.}
    \label{fig:hep2d_eff_true}
\end{figure}

Fig.~\ref{fig:hep2d_eff_true} shows the efficiency for $\nu_{1}, \nu_{2,3} \in [0.5, 1.5]$, normalized by the central value, $\epsilon(1, 1)$. These input points and their corresponding normalized efficiency values are used 
as the ground truth for evaluating the models employed throughout this section.

Directly evaluating the efficiency can be an expensive task, particularly when it involves datasets that are large in size and dimensionality. In such cases, simple regression techniques are often employed. These techniques often assume that efficiency can be factorized in terms of the individual nuisance parameters. As an example of such techniques, we consider the following regression, which we call the \textit{on-axis regression} (OAR):
\begin{align}
    \epsilon(\nu_{1}, \nu_{2,3})_{\mathrm{OAR}} \cong \epsilon(\nu_{1}, 1) \times \epsilon(1, \nu_{2,3}).
    \label{eq:eff_approx}
\end{align}
The idea behind this simple technique is to sample the efficiency along the central axes of the input space, which in this case correspond to $\nu_{1}, \nu_{2,3} = 1$. These samples are then used to estimate the rest of the efficiency by assuming that the \textit{off-axis} values can be approximated by the product of the corresponding \textit{on-axis} values. 

\begin{figure}[!htb]
    \centering
    \includegraphics[width=0.7\textwidth]{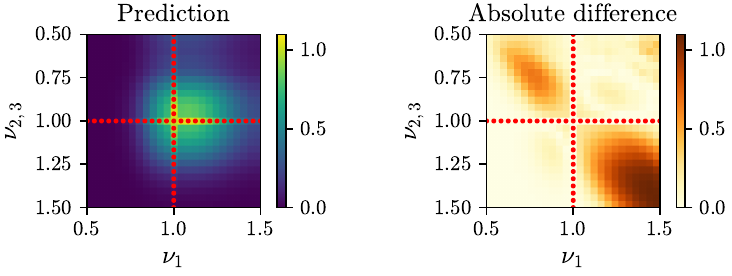}
    \caption{Prediction (left) of the normalized efficiency by on-axis regression, and absolute difference (right) between ground truth and prediction. The red dots indicate the input locations of the samples used in the on-axis regression.}
    \label{fig:hep2d_onaxis}
\end{figure}

To implement the on-axis regression, we sample 49 on-axis observations (24 along each axis plus the central value), which are used to estimate the off-axis efficiency according to Eq.~\ref{eq:eff_approx}. The results are shown in Fig.~\ref{fig:hep2d_onaxis}. The on-axis regression is significantly cheaper than evaluating the efficiency at every input location, but the performance is poor as it fails to capture correlations between the off-axis variables. For example, Fig.~\ref{fig:hep2d_eff_true} shows that the highest efficiency values are concentrated near the diagonal starting from the top left corner. The on-axis regression fails to predict this diagonal correctly. Instead, it predicts the center of the input space to be the area with the highest efficiency. 

\subsection{\label{sec:HEP2d_GP} Gaussian Process Regression}

\begin{figure}[!htb]
    \centering
    \begin{subfigure}{\textwidth}
        \centering
        \includegraphics[width=0.85\linewidth]{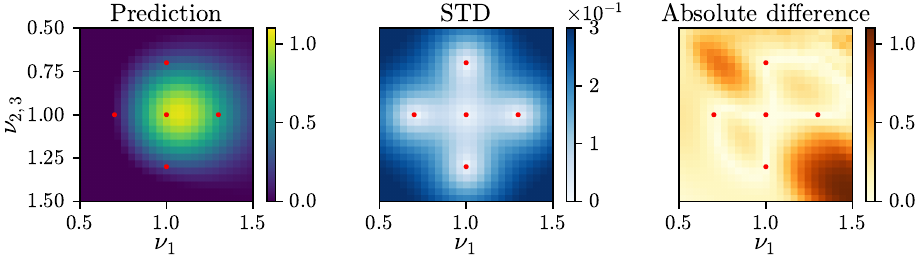}
        \caption{Regular GP}
        \label{fig:hep2d_gp_reg}
    \end{subfigure}

    \begin{subfigure}{\textwidth}
        \centering
        \includegraphics[width=0.85\linewidth]{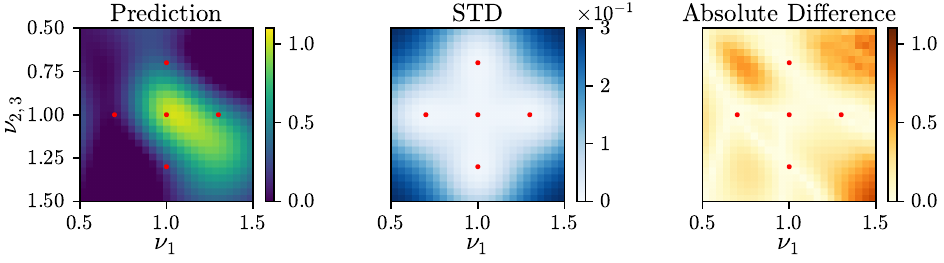}
        \caption{Derivative GP}
        \label{fig:hep2d_gp_der}
    \end{subfigure}
    \caption{Prediction (left), standard deviation (middle), and absolute difference (right) between ground truth and prediction by the specified GP model. The red dots represent the samples used to fit the model.}
\end{figure}

Next, we evaluate the use of GP regression with and without derivative information to approximate the efficiency. Like in the on-axis regression, we begin by sampling the efficiency along the $\nu_{1}, \nu_{2,3} = 1$ axes. To test the predictive power of the GPs, we only sample five observations (two along each axis plus the central value):
\begin{equation*}
    (\nu_{1}, \nu_{2,3}) = (0.7, 1.0), (1.0, 1.0), (1.3, 1.0),(1.0, 0.7), (1.0, 1.3). 
\end{equation*}
These five samples are used to condition a regular GP model with hyperparameters $\alpha=\sqrt{0.1}$, $\ell=0.25$, and $\sigma=0.01$. These hyperparameters are selected by grid search. The results are shown in Fig.~\ref{fig:hep2d_gp_reg}. The prediction is similar to that of the on-axis regression. Both predict higher efficiencies near the center of the grid and have the highest error along the diagonal. 


We also condition a derivative GP model using the same five training samples, along with their gradients. To obtain the gradients, we replace the inequalities in~\cref{eq:sc,eq:effcut} with sigmoid functions. Each sigmoid is shifted to match the 200 GeV threshold ($c=200$) and scaled by $\mid a \mid = 1/10$ to smooth out the gradients. The sign of $a$ is determined by the sign of the inequality; see eq.~\ref{eq:sigmoid}.

For consistency, the derivative GP model uses the same hyperparameters as the regular GP model, in addition to the gradient noise parameter $\sigma_{\mathrm{der}}=0.1$. The results are shown in Fig.~\ref{fig:hep2d_gp_der}. The derivative GP does surprisingly well; with only five samples, it is able to capture the general trend along the diagonal. 


\subsection{\label{sec:HEP2d_BED} Bayesian Experimental Design}

\begin{figure}[!htb]
    \centering
    
    \begin{subfigure}{.78\textwidth}
      \centering
      \includegraphics[width=\linewidth]{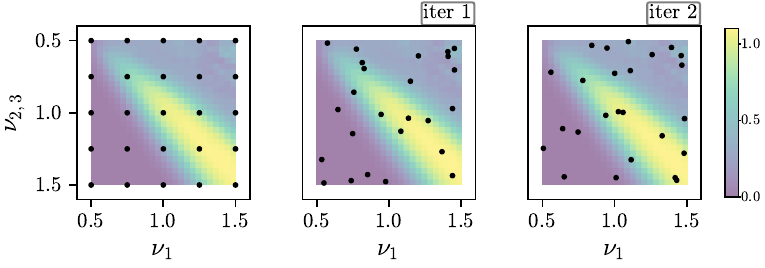}
      \caption{$X_u: 5 \times 5$ grid with $\gamma^2 = 1/10$}
    \end{subfigure}%
    
    \begin{subfigure}{.78\textwidth}
      \centering
      \includegraphics[width=\linewidth]{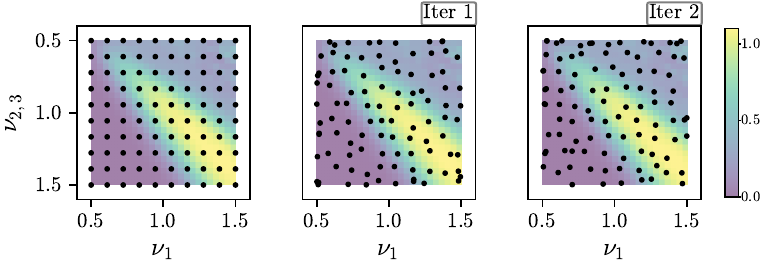}
      \caption{$X_u: 10 \times 10$ grid with $\gamma^2 = 1/50$}
    \end{subfigure}%

\caption{The top panel (a) shows samples of the choice of $5 \times 5$ utility input before (left) and after adding Gaussian noise with variance $\gamma^2=1/10$ (two right-most images). Similarly, the bottom panel (b) shows samples of the choice of $10 \times 10$ utility input with added Gaussian noise with variance $\gamma^2=1/50$. The color scale of the jet efficiency shown in the background is dimmed down for clarity.}
\label{fig:X_u_noisy}
\end{figure}

Finally, we evaluate the use of the BED strategy by executing 44 iterations of the BED algorithm, leading to a total of 49 training samples, the same number used in the on-axis regression strategy. In particular, we examine the sensitivity {of the performance} of the BED strategy with respect to the number of grid points in the utility input $X_u$ and the regularization parameter $\gamma$. We consider the following scenarios: a $25 \times 25$ grid with $\gamma^2 = 1/125$ and $\gamma^2 = 0$; a $10 \times 10$ grid with $\gamma^2 = 1/50$ and $\gamma^2 = 0$; and a $5 \times 5$ grid with $\gamma^2 = 1/10$ and $\gamma^2 = 0$. Two example grid configurations are shown in Fig.~\ref{fig:X_u_noisy}.

\begin{figure}[!htb]
    \centering
    
    \begin{subfigure}{.85\textwidth}
      \centering
      \includegraphics[width=\linewidth]{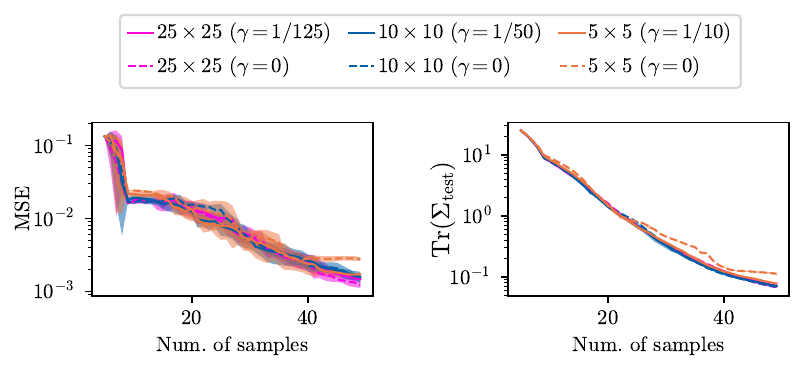}
      \caption{Regular GP}
      \label{fig:hep2d_regGP_Xu}
    \end{subfigure}%
    
    \begin{subfigure}{.85\textwidth}
      \centering
      \includegraphics[width=\linewidth]{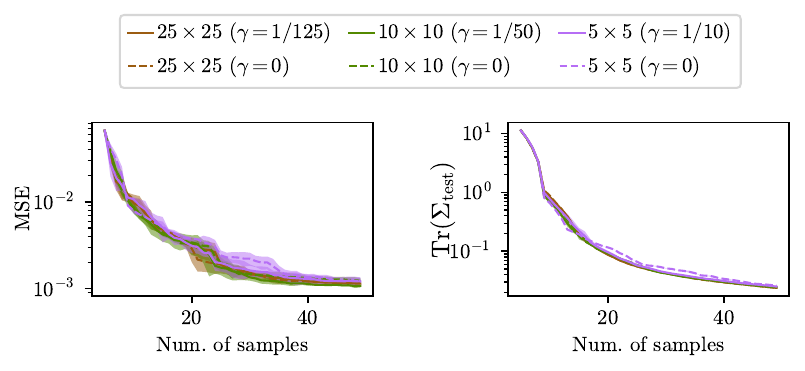}
      \caption{Derivative GP}
      \label{fig:hep2d_derGP_Xu}
    \end{subfigure}%

    \begin{subfigure}{.85\textwidth}
        \centering
        \includegraphics[width=\textwidth]{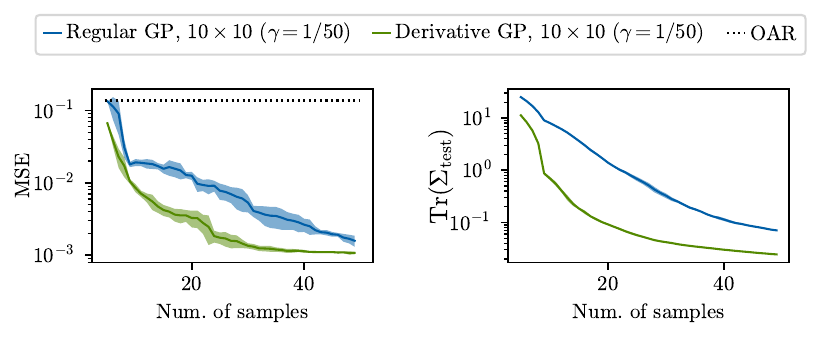}
        \caption{Comparison.}
        \label{fig:hep2d_reg_vs_der}
    \end{subfigure}%

\caption{Panel (a) shows the MSE and $\mathrm{Tr}(\Sigma_{\mathrm{test}})$ of the regular GP model after each of the 44 BED iterations with the various choices of utility input ($X_u$). Similarly, panel (b) shows the MSE and $\mathrm{Tr}(\Sigma_{\mathrm{test}})$ of the derivative GP model. The models are compared in panel (c), along with the MSE of the on-axis regression.  For simplicity, panel (c) only shows the results of the GP models with the choice of $X_u$ of $10 \times 10$ with regularization. The solid lines represent the mean values of 10 calls of the BED strategy with different initial random seeds, and the shaded areas represent the standard deviation.}
\label{fig:hep2d_all_Xu}
\end{figure}

After each BED iteration, the models are evaluated on the test set, a fine $25 \times 25$ grid uniformly spanning the range $\nu_1, \nu_{2,3} \in [0.5, 1.5]$, to update the predictive mean and covariance matrix. From these quantities, we compute the MSE and $\mathrm{Tr}({\Sigma}_{\mathrm{test}})$ values. The results are shown in Figs.~\ref{fig:hep2d_regGP_Xu} and ~\ref{fig:hep2d_derGP_Xu}.  

Overall, we observe a trade-off between the resolution of the grids and the efficacy of the BED strategy: higher-resolution grids result in lower overall MSE and $\mathrm{Tr}({\Sigma}_{\mathrm{test}})$ values. However, this trade-off is only marginal when regularization is applied. For instance, the choice of $5 \times 5$ utility input without regularization yields the highest overall MSE and $\mathrm{Tr}({\Sigma}_{\mathrm{test}})$ values across both the regular and derivative GP models. With regularization, the performance of the models improves significantly, nearly matching that of models paired with more expensive, finer grid choices.

The results of the regular and derivative GP models are compared in Fig.~\ref{fig:hep2d_reg_vs_der}. For clarity, we only show the results of the BED strategy with the choice of $10 \times 10$ utility input with regularization. The derivative GP consistently outperforms the regular GP with lower MSE and $\mathrm{Tr}({\Sigma}_{\mathrm{test}})$ values.

The MSE of the on-axis regression is shown in Fig.~\ref{fig:hep2d_reg_vs_der}. This error is initially similar to that of the regular GP model, and nearly twice as large as that of the derivative GP model. As additional training inputs are selected using the BED strategy, the MSE and $\mathrm{Tr}({\Sigma}_{\mathrm{test}})$ values of the GP models decrease rapidly. By the time the BED has selected the same number of samples as the on-axis regression, the error of the GP models is significantly smaller. These results highlight the capabilities of GP-based methods for estimating detector observables, particularly when compared to simpler regression techniques that assume factorization of the nuisance parameters.

\begin{figure}[!htb]
    \centering
    
    \begin{subfigure}{.5\textwidth}
      \centering
      \includegraphics[width=0.93\linewidth]{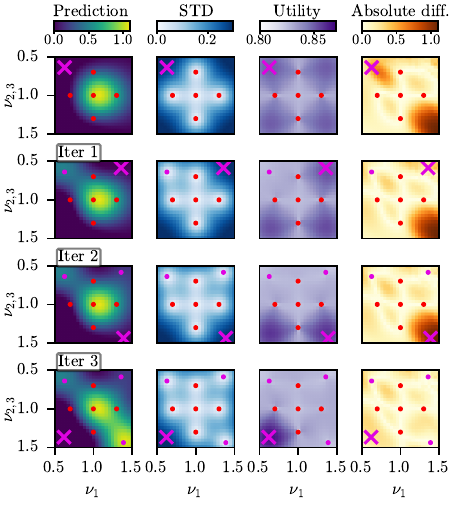}
      \caption{Regular GP}
      \label{fig:hep2d_bed_reg}
    \end{subfigure}%
    \begin{subfigure}{.5\textwidth}
      \centering
      \includegraphics[width=0.93\linewidth]{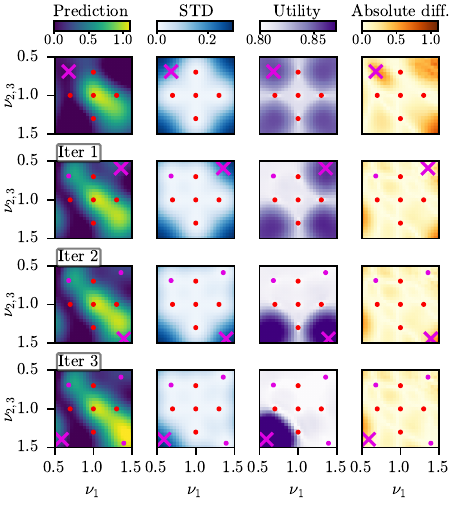}
      \caption{Derivative GP}
      \label{fig:hep2d_bed_der}
    \end{subfigure}%

\caption{Prediction (first column), standard deviation (second column), utility (third column), and absolute difference (fourth column) between ground truth and prediction by the specified GP model. The panels show the initial training samples (red dots), the training samples selected by the BED (pink dots), and the location of the next sample selected by the BED by maximizing the utility (pink cross).}
\label{fig:hep2d_bed_all}
\end{figure}

A visualization of the first three BED iterations is shown in Fig.~\ref{fig:hep2d_bed_all}. With every iteration, the utility is maximized and a new training sample is selected. As expected, the images show that once the models are reconditioned on a new sample, the utility and standard deviation (STD) near it are reduced.

The goal of the BED strategy is to choose samples that efficiently reduce the predictive variance of a model. A critical question is whether it achieves this goal in practice. To answer this question, we compare the BED strategy with two alternative sampling methods: random and grid sampling. In the random sampling case, we sequentially add 44 random observations to the training set. In the grid sampling case, we sample squared grids of size $n \times n$, $n=2, \dots, 6$, each combined with the initial five training samples. Unlike the BED and random sampling strategies, the grid sampling strategy is not sequential. Examples of the models when conditioned using random and grid sampling methods are provided in Appendices~\ref{app:random} and~\ref{app:grid}, respectively.

\begin{figure}[!htb]
    \centering
    
    \begin{subfigure}{.85\textwidth}
      \centering
      \includegraphics[width=\linewidth]{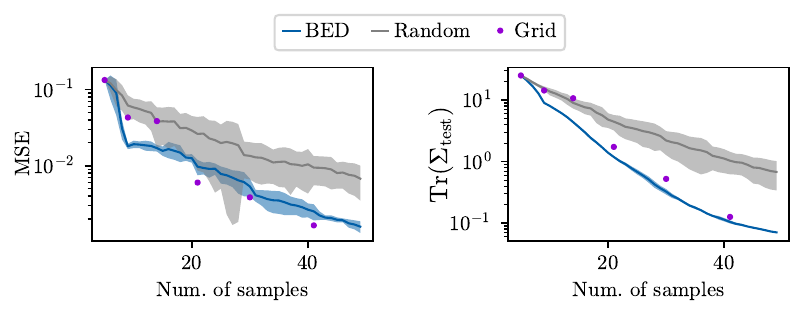}
      \caption{Regular GP}
      \label{fig:hep2d_BED_mse_uncert_regGP}
    \end{subfigure}%
    
    \begin{subfigure}{.85\textwidth}
      \centering
      \includegraphics[width=\linewidth]{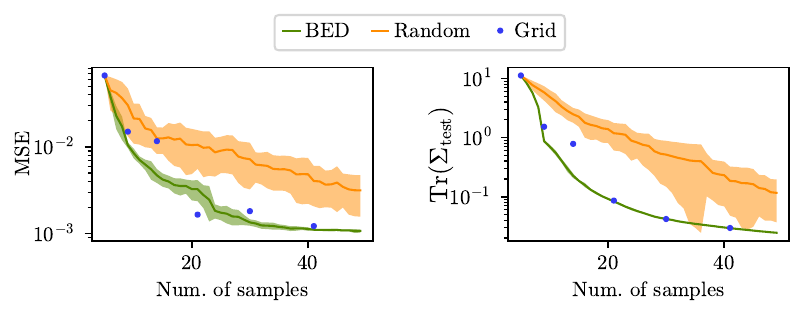}
      \caption{Derivative GP}
      \label{fig:hep2d_BED_mse_uncert_derGP}
    \end{subfigure}%

\caption{MSE and $\mathrm{Tr}(\Sigma_{\mathrm{test}})$ after each iteration of the various sampling strategies. In the BED and random sampling cases, the solid lines represent the mean values of 10 calls of the sampling strategies using different initial random seeds, and the shaded areas represent the standard deviation.}
\label{fig:hep2d_BED_mse_uncert_all}
\end{figure}

The performance of the various sampling strategies is shown in Fig.~\ref{fig:hep2d_BED_mse_uncert_all}. For clarity, we only show the results of the GP models with the choice of utility input of $10 \times 10$ with 
 regularization, but the results generalize to the other choices. Random sampling performs the worst, with the highest overall MSE and $\mathrm{Tr}(\Sigma_{\mathrm{test}})$ values. Grid sampling performs better, {with results comparable to those of the BED strategy}. The BED achieves the lowest overall $\mathrm{Tr}(\Sigma_{\mathrm{test}})$ values, demonstrating its effectiveness in reducing the predictive uncertainty.

In this example, grid sampling performs similarly to the BED strategy. However, in real-world examples, the efficiency is likely to depend on multiple nuisance parameters, which create complex, high-dimensional surfaces that must be estimated. Sampling such spaces with uniform grids is generally suboptimal because the size of the grids grows exponentially with the dimensionality of the space. In the following section, we continue our comparison of the various sampling strategies in a higher-dimensional space (4D), where the advantages of a Bayesian-based strategy may become more evident.

\section{\label{sec:HEP_4D}High-Energy Physics Experiments: 4D Efficiency Estimation}

\begin{figure}[!htb]
    \centering
    \includegraphics[width=0.38\textwidth]{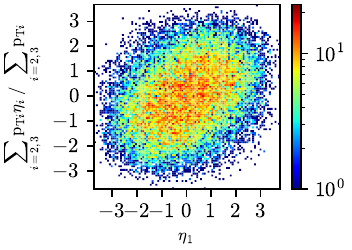}
    \caption{Distribution of the pseudorapidity of $j_1$, and the $\pT$-weighted pseudorapidity mean of $j_2$ and $j_3$.}
    \label{fig:hep4d_etadist}
\end{figure}

In this Section, we increase the number of nuisance parameters by considering four jet energy scales, which are dependent on the pseudorapidity ($\eta$) of the jets, and roughly correspond to the central and outer calorimeter regions:
\begin{itemize}
    \item $\nu_{1}^{\mathrm{central}}$: The jet energy scale of $j_1$ in the {region where its pseudorapidity is less than one:} $\mid \eta_1 \mid < 1$.
    \item $\nu_{1}^{\mathrm{outer}}$: The jet energy scale of $j_1$ {elsewhere: }$\mid \eta_1 \mid \geq 1$.
    \item $\nu_{2,3}^{\mathrm{central}}$: The jet energy scale of $j_2$ and $j_3$ in the {region where their $\pT$-weighted mean pseudorapidity is less than one: }\\ $ \mid \sum\limits_{i=2, 3} \pT_i \eta_i / \sum\limits_{i=2, 3} \pT_i \mid < 1$.
    \item $\nu_{2,3}^{\mathrm{outer}}$: The jet energy scale of $j_2$ and $j_3$ in the {elsewhere: }{} \\ $ \mid \sum\limits_{i=2, 3} \pT_i \eta_i / \sum\limits_{i=2, 3} \pT_i \mid \geq  1$.
    \label{eq:sc_eta}
\end{itemize}
Where $\eta_i$ is the pseudorapidity of $j_i$. The distribution of the pseudorapidities is shown in Fig.~\ref{fig:hep4d_etadist}. For every event, either $\nu_{1}^{\mathrm{central}}$ or $\nu_{1}^{\mathrm{outer}}$ is selected as the jet energy scale of $j_1$. Likewise, either $\nu_{2,3}^{\mathrm{central}}$ or $\nu_{2,3}^{\mathrm{outer}}$ is selected as the jet energy scale of $j_2$ and $j_3$. The efficiency is then calculated according to the procedure specified in Sec.~\ref{sec:HEP_2D} and normalized by the central value, $\epsilon(1, 1, 1, 1)$. 

\begin{figure}[!htb]
    \centering
    \includegraphics[width=0.38\textwidth]{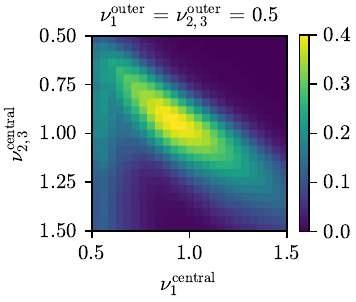}
    \caption{Normalized efficiency as $\nu_{1}^{\mathrm{central}}$ and $\nu_{2,3}^{\mathrm{central}}$ are varied in the range $[0.5, 1.5]$ while $\nu_{1}^{\mathrm{outer}}$=$\nu_{2,3}^{\mathrm{outer}} = 0.5$.}
    \label{fig:hep4d_truth}
\end{figure}

Visualizing the efficiency as a function of the four jet energy scales is challenging due to the high dimensionality of the input space. For illustration purposes, we show only a slice of the normalized efficiency by plotting it as we vary $\nu_{1}^{\mathrm{central}}$ and $\nu_{2,3}^{\mathrm{central}}$ in the range $[0.5, 1.5]$ while we fix $\nu_{1}^{\mathrm{outer}}$ and $\nu_{2,3}^{\mathrm{outer}}$ to have a magnitude of $0.5$. The selected slice of the normalized efficiency is shown in Fig.~\ref{fig:hep4d_truth}.

In this section, we test all models on a fine grid of size $25^4$, uniformly distributed in the range $\nu_{1}^{\mathrm{central}}, \nu_{1}^{\mathrm{outer}}, \nu_{2,3}^{\mathrm{central}}, \nu_{2,3}^{\mathrm{outer}} \in [0.5, 1.5]$. As a benchmark, and to visualize the quality of the models, we show the models' predictions on the selected slice of the normalized efficiency shown in Fig.~\ref{fig:hep4d_truth}.

\begin{figure}[!htb]
    \centering
    \includegraphics[width=0.7\textwidth]{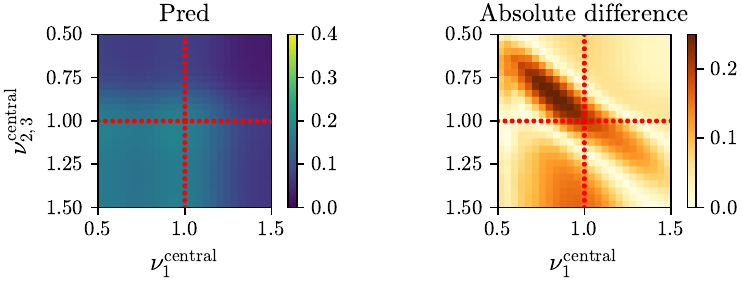}
    \caption{Prediction (left) of the selected slice of the normalized efficiency by on-axis regression and the absolute difference (right) between ground truth and prediction. The red dots represent the samples used in the on-axis regression.}
    \label{fig:hep4d_onaxis}
\end{figure}

We approximate the efficiency by on-axis regression. A total of 97 on-axis observations (24 along each central of the four axes plus the central value) are sampled and used to estimate the off-axis values according to
\begin{equation}
    \epsilon (\nu_{1}^{\mathrm{central}}, \nu_{1}^{\mathrm{outer}}, \nu_{2,3}^{\mathrm{central}}, \nu_{2,3}^{\mathrm{outer}})_{\mathrm{OAR}} \cong 
    \epsilon (\nu_{1}^{\mathrm{central}}, 1, 1, 1) \times 
    \epsilon (1, \nu_{1}^{\mathrm{outer}}, 1, 1)\times 
    \epsilon (1, 1, \nu_{2,3}^{\mathrm{central}}, 1) \times
    \epsilon (1, 1, 1, \nu_{2,3}^{\mathrm{outer}}).
\end{equation}
Fig.~\ref{fig:hep4d_onaxis} shows the results of the on-axis regression on the selected slice of the normalized efficiency. This simple regression is cheap but performs poorly as it again fails to predict the higher efficiency values along the diagonal. 

\subsection{\label{sec:HEP4d_GP} Gaussian Process Regression}

Next, we evaluate the use of GP regression with and without derivative information to approximate the efficiency. We sample nine observations along the central axes (two along each of the four axes plus the central value). Like in the 2D case, these initial observations are placed on-axis at $0.7$, $1$, and $1.3$. Using these nine observations, we train both a regular and a derivative GP model with the same hyperparameters used in the 2D case\footnote{We quickly tested other hyperparameters, but the ones selected in the 2D case also worked well enough for the 4D case.}. 

To obtain the gradients with respect to all four jet energy scales, we replace the pseudorapidity-based cuts with sigmoid functions. These sigmoids are shifted by $c=1$ to match the $\eta$ threshold, and scaled by $\mid a \mid = 1/10$ to smooth out the gradients. The sign of $a$ is determined by the sign of the inequality; see Eq.~\ref{eq:sigmoid}. The gradient noise parameter is set to $\sigma_{\mathrm{der}}=0.1$. 

\begin{figure}[!htb]
    \centering
    
    \begin{subfigure}{.85\textwidth}
      \centering
      \includegraphics[width=\linewidth]{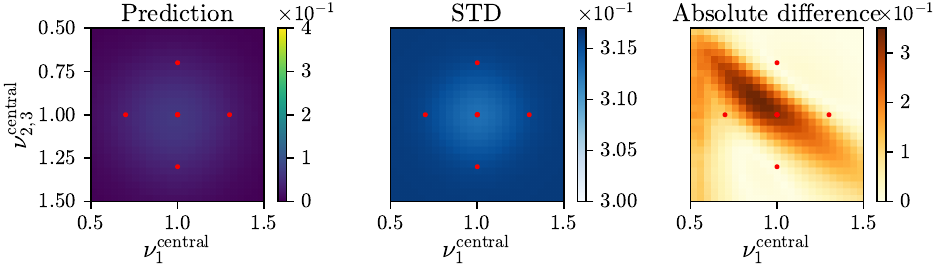}
      \caption{Regular GP}
    \end{subfigure}%
    
    \begin{subfigure}{.85\textwidth}
      \centering
      \includegraphics[width=\linewidth]{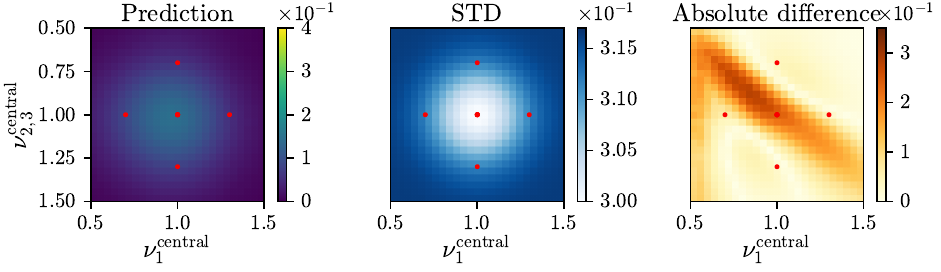}
      \caption{Derivative GP}
    \end{subfigure}%

\caption{Prediction (left) of the selected slice of the normalized efficiency, standard deviation (middle), and the absolute difference (right) between ground truth and prediction by the specified GP model. The red dots represent the samples used to train the GP models.}
\label{fig:hep2d_gpr}
\end{figure}

The predictions made by the GP models are shown in Fig.~\ref{fig:hep2d_gpr}. Both models predict higher efficiency values near the center of the slice. Unlike in the 2D scenario, the advantages of the derivative GP model are not as evident with the initial training set.

\subsection{\label{sec:HEP4d_BED}Bayesian Experimental Design}

Finally, we evaluate the use of the BED strategy by executing 88 iterations of the BED algorithm, leading to a total of 97 training samples, the same number used in the on-axis regression strategy. For the utility input, $X_{u}$, we consider 4D grids uniform (of size $n^4$) of the following sizes, with and without regularization\footnote{Higher resolution grids are not considered due to high computational costs.}: $n=10$ with $\gamma^2 = 1/50$ and $\gamma^2 = 0$; $n=5$ with $\gamma^2 = 1/10$ and $\gamma^2 = 0$; and $n=3$ with $\gamma^2 = 1/5$ and $\gamma^2 = 0$.

\begin{figure}[!htp]
    \centering
    
    \begin{subfigure}{.8\textwidth}
      \centering
      \includegraphics[width=\linewidth]{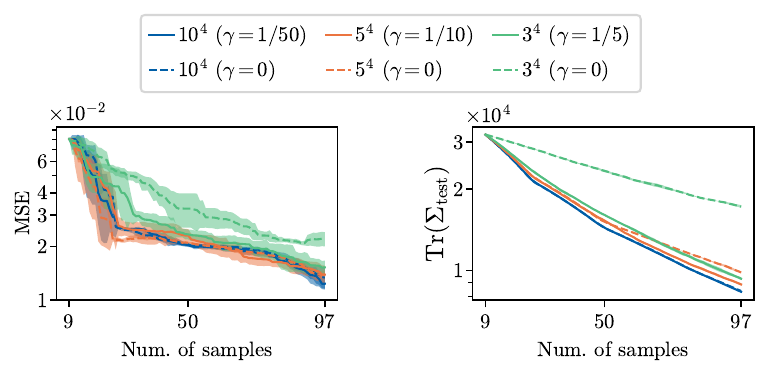}
      \caption{Regular GP}
      \label{fig:hep4d_regGP_Xu}
    \end{subfigure}%
    
    \begin{subfigure}{.8\textwidth}
      \centering
      \includegraphics[width=\linewidth]{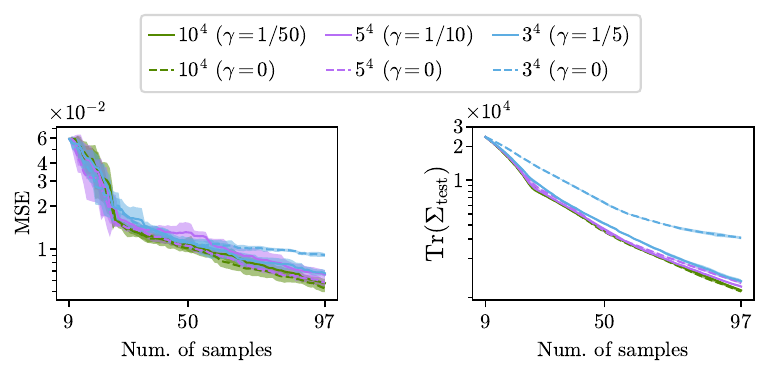}
      \caption{Derivative GP}
      \label{fig:hep4d_derGP_Xu}
    \end{subfigure}%

    \begin{subfigure}{.8\textwidth}
      \centering
      \includegraphics[width=\linewidth]{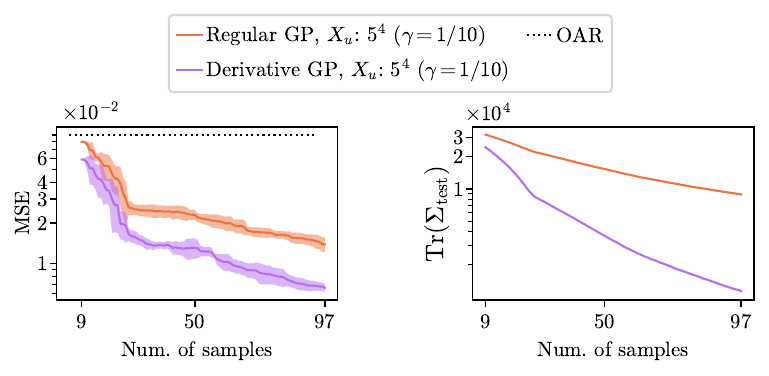}
      \caption{Comparison}
      \label{fig:hep4d_reg_vs_der}
    \end{subfigure}%
    
\caption{Panel (a) shows the MSE and $\mathrm{Tr}(\Sigma_{\mathrm{test}})$ of the regular GP model after each of the 88 BED iterations with the various choices of utility input ($X_u$). Similarly, panel (b) shows the MSE and $\mathrm{Tr}(\Sigma_{\mathrm{test}})$ of the derivative GP model, and panel (c) shows their comparison along with the MSE of the on-axis regression. For clarity, panel (c) only shows the results of the GP models with the choice of n=5 utility input with $\gamma^2=1/10$ regularization. The solid lines represent the mean values of five calls of the BED strategy with different initial random seeds, and the shaded areas represent the standard deviation.}
\label{fig:hep4d_all_Xu}
\end{figure}

After each iteration, the models are evaluated on a high-resolution grid with $n=25$, i.e. $25^4$ points. The grid points are uniformly distributed in the range $\nu_{1}^{\mathrm{central}}$, $\nu_{1}^{\mathrm{outer}}$, $\nu_{2,3}^{\mathrm{central}}$, and $\nu_{2,3}^{\mathrm{outer}} \in [0.5, 1.5]$.
The results are shown in Fig.~\ref{fig:hep4d_all_Xu}. 

The benefit of adding regularization to the utility input is more evident than in the 2D case. The sparsest choice of utility input we consider is a regular grid with $n=3$ (i.e. $3^4$ grid points). Without regularization, this $n=3$ grid gives the highest MSE and $\mathrm{Tr}(\Sigma_{\mathrm{test}})$ values by and large, for both regular and derivative GP models. The performance of the corresponding models improves significantly when regularization is added, achieving comparable results to the models paired with the more costly grid choices. These results are important because the cost of inference grows exponentially with the dimensionality of the input space, and thus, having an economical yet efficient sampling strategy can be crucial.

Fig.~\ref{fig:hep4d_reg_vs_der} compares the performance of the regular and derivative GP models, as well as the MSE of the on-axis regression. For simplicity, we only show the results of the BED strategy with the choice of $n=5$ utility input with $\gamma=1/10$ regularization. Initially, the regular and derivative GP models have similar MSE and $\mathrm{Tr}(\Sigma_{\mathrm{test}})$ values. As additional training samples are incorporated, the derivative GP begins to outperform the regular GP. Both GP models have significantly lower MSE values than the on-axis regression.

\begin{figure}[!ht]
    \centering
    
    \begin{subfigure}{.8\textwidth}
      \centering
      \includegraphics[width=\linewidth]{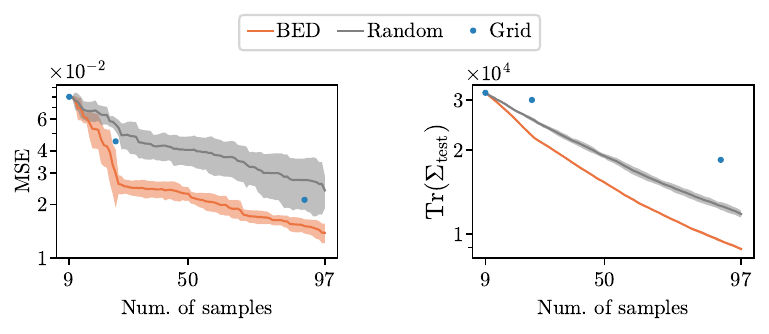}
      \caption{Regular GP}
      \label{fig:hep4d_BED_mse_uncert_regGP}
    \end{subfigure}%
    
    \begin{subfigure}{.8\textwidth}
      \centering
      \includegraphics[width=\linewidth]{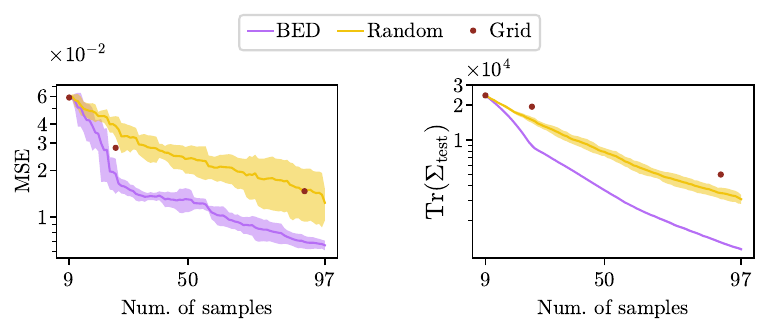}
      \caption{Derivative GP}
      \label{fig:hep4d_BED_mse_uncert_derGP}
    \end{subfigure}%

\caption{MSE and $\mathrm{Tr}(\Sigma_{\mathrm{test}})$ after each iteration of the various sampling strategies. In the BED and random sampling cases, the solid lines represent the mean values of five calls of the sampling strategies using different initial random seeds and the shaded areas the standard deviation.}
\label{fig:hep4d_BED_mse_uncert_all}
\end{figure}

To answer the critical question of whether the BED strategy has efficiently reduced the variance of the model, we again compare the results to random and grid sampling. In the random sampling case, we sequentially add 88 uniformly distributed random observations to the training set. In the grid sampling case, we sample 4D uniform grids of size $n=2$ and $n=3$, each combined with the initial training set. The results are shown in Fig.~\ref{fig:hep4d_BED_mse_uncert_all}.

The results clearly demonstrate that the BED strategy outperforms the other sampling methods. Its advantages become more evident in high-dimensional spaces, where sampling is notoriously difficult due to the curse of dimensionality. The grid sampling strategy performs the worst. While uniform grids are common when sampling low-dimensional spaces, they are rarely optimal in higher dimensions because of their exponential scaling. Random sampling performs better, but it has large variations in the MSE. Overall, the BED strategy has the lowest MSE and $\mathrm{Tr}(\Sigma_{\mathrm{test}})$ values, regardless of the choice of utility input or GP model.

\section{\label{sec:concl} Conclusions}

Physics experiments, like those conducted at the Large Hadron Collider, play a crucial role in precision tests of the Standard Model, where experimental measures of particle properties are compared to theoretical predictions. A critical aspect of these tests is the accurate assessment of systematic uncertainties. 
%


{Ideally, after identifying the sources of uncertainty that affect an analysis, one would thoroughly and quantitatively map out the effects of those uncertainties on the experimental response. It is typically straightforward to associate nuisance parameters to the sources of uncertainty; however, estimating the response as a function of the nuisance parameters is challenging -- it corresponds to estimating a function of several variables and is subject to the curse of dimensionality. }

{Several techniques have been introduced to estimate the effect of multiple sources of systematic uncertainties. Most assume the individual sources of uncertainty are independent, though it is fairly straightforward to incorporate correlations. More critically, most approaches also assume that the effect of those uncertainties factorizes, which is much less motivated.  In this paper, we provide examples where these assumptions fail and propose a method for estimating the joint response to multiple sources of systematic uncertainties without assuming they factorize.} Our approach uses Gaussian process regression to model observables of the experimental response as a function of the nuisance parameters. 
{When (approximate) gradients of the response with respect to the nuisance parameters can be computed, that informaiton can also be incorporated into the regression.}
{We show that GPs enhanced with derivative information can efficiently estimate the complex and unfactorizable effects of multiple sources of systematic uncertainty.}

{We evaluate our approach in two high-energy physics scenarios, calculating the efficiency of events as a function of two and four nuisance parameters. In all scenarios, the GP regression trained with gradients consistently outperforms the regular GP regression, even when trained on limited samples. Both models significantly surpass the performance of a more traditional approach assuming the factorization of the nuisance parameters.}

{We also introduce a sampling strategy based on Bayesian experimental design (BED). This sampling strategy is designed to select new observations that efficiently reduce a GP model's predictive uncertainty. In the examples presented in this paper, we show that the BED sampling strategy is more efficient at reducing the predictive variance of the GP models than either grid or random sampling.}

\ack
K.C. was supported by NSF PHY-2323298. A.R. and D.W. are supported by the DOE Office of Science.




\suppdata
Supplementary material includes the installable package ``gpder''. The source code, along with detailed demos of how to use the package, can be found at https://github.com/alexxromero/gpder.

\appendix
\section{Random Sampling}
\label{app:random}

\begin{figure*}[!ht]
    \centering
    
    \begin{subfigure}{.5\textwidth}
      \centering
      \includegraphics[width=0.95\linewidth]{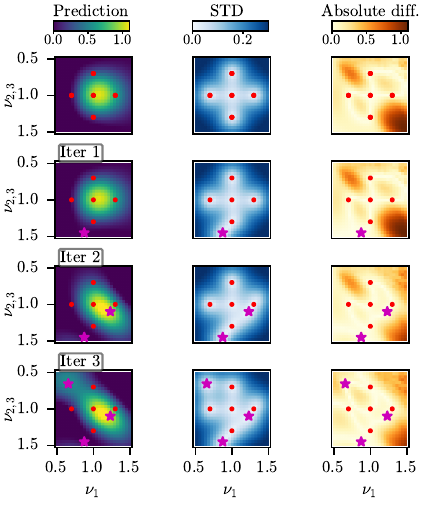}
      \caption{Regular GP}
    \end{subfigure}%
    \begin{subfigure}{.5\textwidth}
      \centering
      \includegraphics[width=0.95\linewidth]{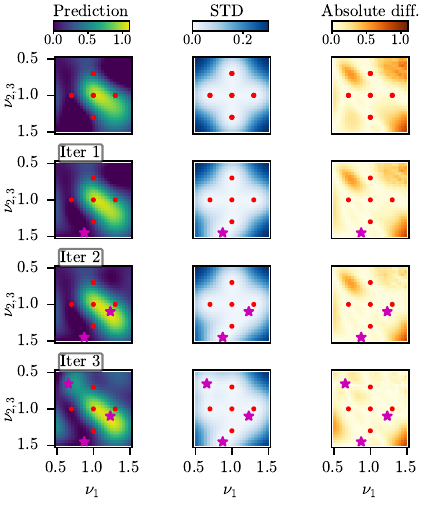}
      \caption{Derivative GP}
    \end{subfigure}%

\caption{Prediction (first column), standard deviation (second column), utility (third column), and absolute difference (fourth column) between ground truth and prediction by the specified GP model. The panels show the initial training samples (red dots) and the random training samples (purple stars).}
\label{fig:hep2d_random}
\end{figure*}

In this appendix, we illustrate the random sampling strategy from Sec.~\ref{sec:HEP_2D}. The evolution of the GP models after three iterations of the random sampling strategy is shown in Fig.~\ref{fig:hep2d_random}. After each iteration, a new random sample is added to the training set.

\section{Grid Sampling}
\label{app:grid}

\begin{figure*}[!ht]
    \centering
    
    \begin{subfigure}{.5\textwidth}
      \centering
      \includegraphics[width=0.95\linewidth]{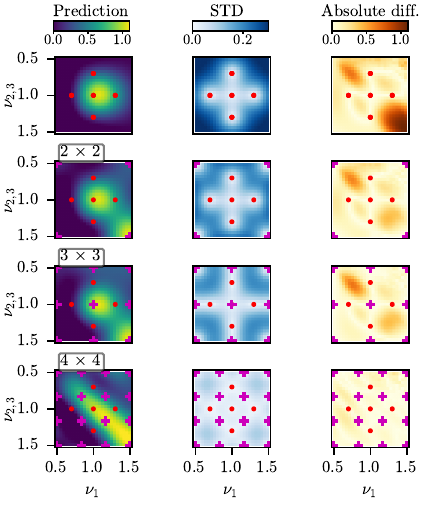}
      \caption{Regular GP}
    \end{subfigure}%
    \begin{subfigure}{.5\textwidth}
      \centering
      \includegraphics[width=0.95\linewidth]{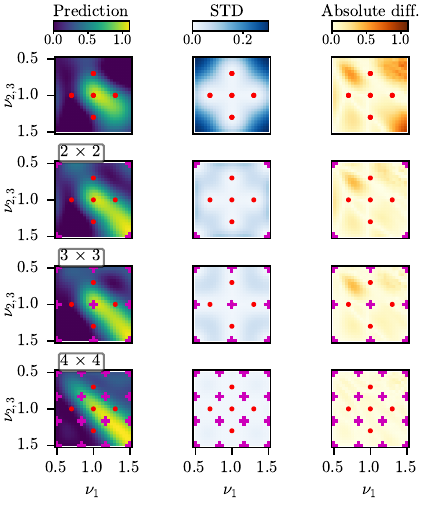}
      \caption{Derivative GP}
    \end{subfigure}%

\caption{Prediction (first column), standard deviation (second column), utility (third column), and absolute difference (fourth column) between ground truth and prediction by the specified GP model. The panels show the initial training samples (red dots) and the grid training samples (purple plus signs).}
\label{fig:hep2d_grid}
\end{figure*}

In this appendix, we illustrate the grid sampling strategy from Sec.~\ref{sec:HEP_2D}. The evolution of the GP models after the addition of $n \times n$, $n=2, 3$, or $4$ grids to the training set is shown in Fig.~\ref{fig:hep2d_grid}.

\bibliographystyle{iopart-num}
\bibliography{references}

\end{document}